\documentclass[10pt, twocolumn]{article}
\usepackage[a4paper, margin=1.5cm]{geometry}
\usepackage[utf8]{inputenc} 
\usepackage[T1]{fontenc}    
\usepackage{hyperref}       
\usepackage{url}   
\usepackage{amsmath,amssymb,amsfonts}
\usepackage{algorithmic}
\usepackage{graphicx}
\usepackage{textcomp}
\usepackage{ctable}
\usepackage{subcaption}
\usepackage{wrapfig}
\usepackage{multirow}
\usepackage{dsfont}
\usepackage{breakcites}
\usepackage{soulutf8}
\usepackage{xcolor}
\usepackage{url}
\usepackage{amsmath}
\usepackage{caption}
\usepackage{multirow}

\begin{document}
\title{ `Just One More Sensor is Enough' -- Iterative Water Leak Localization with Physical Simulation and a Small Number of Pressure Sensors}
\author{Michał Cholewa, Michał Romaszewski, Przemysław Głomb, \\ Katarzyna Kołodziej, Michał Gorawski, Jakub Koral, \\ Wojciech Koral, Andrzej Madej, Kryspin Musioł}
\date{June, 2024\footnote{© 2024 IEEE.  Personal use of this material is permitted.  Permission from IEEE must be obtained for all other uses, in any current or future media, including reprinting/republishing this material for advertising or promotional purposes, creating new collective works, for resale or redistribution to servers or lists, or reuse of any copyrighted component of this work in other works.} }
\maketitle

\begin{abstract}
In this article, we propose an approach to leak localisation in a complex water delivery grid with the use of data from physical simulation (e.g. EPANET software). This task is usually achieved by a network of multiple water pressure sensors and analysis of the so-called sensitivity matrix of pressure differences between the network's simulated data and actual data of the network affected by the leak. However, most algorithms using this approach require a significant number of pressure sensors -- a condition that is not easy to fulfil in the case of many less equipped networks. Therefore, we answer the question of whether leak localisation is possible by utilising very few sensors but having the ability to relocate one of them. Our algorithm is based on physical simulations (EPANET software) and an iterative scheme for mobile sensor relocation. The experiments show that the proposed system can equalise the low number of sensors with adjustments made for their positioning, giving a very good approximation of leak's position both in simulated cases and real-life example taken from BattLeDIM competition L-Town data. 
\end{abstract}

\section{Introduction}

With the growing human population and aridification of many areas of our world, supplying water becomes increasingly challenging. This issue is most visible in countries with limited access to water \cite{neves2010evolvable}, \cite{needs2021progress}. The report prepared for World Bank in 2005 \cite{liemberger2006challenge} indicated that the estimated level of `non-revenue water' (NRW) is 48.6 bln $\text{m}^3$/yr and real leaks from water delivery systems (`real losses') are around 32.7 bln $\text{m}^3$/yr. For the year 2016, Liemberger and Wyatt in \cite{liemberger2019quantifying} showed that the levels of NRW calculated ten years prior were vastly underestimated, and in 2006, NRW levels were significantly higher (97.5 bln $\text{m}^3$/yr), increasing to 126 bln $\text{m}^3$/yr in 2016. At the same time, it was estimated that lowering the level of real losses (the leaks) by 1/3 would lower the yearly costs of water management by 113 bln USD, additionally allowing access to fresh water for another 800 million people (assuming daily usage of 150 litres per person).

These calculations result in NRW reduction being in the field of interest of not only financial entities (World Bank) but also of the industry within the field of water delivery (like European Diehl \cite{diehl}, Hawle \cite{hawle}, AVK \cite{avk} or worldwide Xylem or Myia) as well as tech companies (Bentley, DHI). Following calculations in \cite{liemberger2019quantifying}, one of the most straightforward ways of reducing the amount of NRW is through the development of methods allowing reduction of `real losses' -- which means fast addressing the leaks in water delivery systems (WDS).

Leak detection and localization are complicated issues. According to experts, only 10-20\% of leaks are visible on the ground. The rest has to be detected and localized using different means. That requires a complex and often time-consuming approach choice, which includes the structure and size of the water delivery system, pressure management, and various technical and economic conditions for leak detection methods \cite{hamilton2013leak}. One of the ways to approach this task is from the data-driven standpoint, with analysis of the water balance in the telemetry of District Metered Areas (DMAs), as in~\cite{farah2017leakage},~\cite{yu2021integrated},~\cite{glomb2023detection} or~\cite{jensen2018leakage}. These methods compare the consumption registered by water meters -- data sensors commonly installed for billing purposes, thus often available -- and the inflow to the DMA.

After the leak is detected within the Water Delivery System (WDS), it must be localised. The method most frequently used for precise leak localisation is acoustic method~\cite{KAFLE2022acoustic}, \cite{khulief2012acoustic}, based on analysis of the sound created by water leaking through the rupture. The distance, however, over which the sound of the leak travels along the pipeline is limited to 100--200 meters for steel or cast iron pipelines and 20--100 meters for PE or PCV pipes. Consequently, to use that method for 100 km of pipelines in WDS, over a thousand points (hydrants or gates) need to be analysed, and when using geophones, that number rises to ten thousand, which is costly and time-consuming.

For that reason, the critical element for shortening the time of leak localisation is limiting the area in which it is searched. The most often used method to limit the area is to divide the entire WDS into smaller areas, which can be isolated from the rest of the system. One way to achieve this is by using gates to shut off the water inflow to certain areas and simultaneously analysing flow characteristics. After the area with the leak is isolated, the water flow decreases, thus showing the probable area of the leak. This method has, however, weaknesses -- it effectively cuts off water for the entire isolated area, and there is a risk of a new leak emerging when opening the gates, which will temporarily increase the water pressure. Another way is to divide the WDS into DMAs (District Meter Areas) and install meters on smaller parts of the WDS. Monitoring water inflow to each DMA (especially minimum night flows -- MNF) allows for a faster reaction to the emergence of the leak. This method requires either the division considered in the WDS design or using gates to isolate the DMAs.

Since, for many providers, neither shutting down the entire sectors of WDS nor redesigning the entire network is a valid option, leak localization increasingly often includes the stage of approximate calculations based on the hydraulic model of WDS and pressure data~\cite{ISHIDO2014pressure}. Those are particularly useful for WDS equipped with individual water meters, which allow monitoring water usage in every system fragment and thus get precise information about the water flow. The combination of data from AMI (Automated Metering Infrastructure) and pressure sensors allows for a significant narrowing of the area of the possible leak location. Even in smaller, isolated sub-networks, the problem still needs to be solved. A reliable measurement of pressure changes is required for the algorithm to work accurately \cite{zaman2020review}, and a significant number of sensors relative to the number of nodes in WDS, which is often costly for smaller providers. In \cite{mashhadi2021use}, authors use five sensors for a relatively small network of only 45 pipes and 33 junctions, \cite{li2022leakage} using a network of 491 nodes utilizes 20 pressure sensors, and while \cite{righetti2019optimal} uses only four measure sensors, the used network itself only consists of 34 pipes and 23 users. A Bayesian-based approach is presented in~\cite{SOLDEVILA2017Bayesian} and applied to fragments of the Barcelona municipal network using only six pressure sensors, a very small number of physical sensors in WDS compared to other approaches. Some works try to address this problem; authors of \cite{casillas2013optimal}, \cite{Rastburg2020GamePlacement}, \cite{nejjari2015optimal} try to reduce the number of sensors by choosing their locations within the network, presenting it as an optimization problem - but that reduced number of sensors is often a significant number even in moderately sized zones. 
The practicality, however, presents the WDS administrators with a much more scarce number of sensors. In the authors' experience with more than 20 WDS, a district with approximately 500 nodes can often expect to have one or two pressure sensors installed, which can be a very challenging problem for the leak localization scheme to solve.

A promising alternative lies in the adaptive use of available sensors. Some works have included reallocating the assets at hand to focus on particular instances of the problem arising in WDS. The work of \cite{RICORAMIREZ2007StochasticPlacement} considers a two-stage sensor location algorithm for detecting hazardous substances in the system. The approach includes initial calculations, placing the sensors according to them, calculating the risk for the population and, based on that, modifying the decisions concerning sensor placement.
The idea of moving existing sensors has also been explored in \cite{SOLDEVILA2022Relocation}, where authors propose moving sensors after getting an initial approximation of the detected leak. Both \cite{RICORAMIREZ2007StochasticPlacement} and \cite{SOLDEVILA2022Relocation}, however, still require a significant number of sensors -- both the ones that can be moved and the ones that remain stationary. This article expands on the idea of handling the situation with a smaller number of sensors, which is often the case in many water delivery companies where the infrastructure is older or where economic or administrative conditions make the options for the expansion of pressure networks scarce.

Our experiments on simulated data show that relocation of the single sensor gives a significant improvement and allows the approximation of the leak to be more than twice as accurate in terms of distance. That result can be achieved even in pessimistic situations with only one mobile sensor, and only 33 out of 792 nodes are available for mobile sensor placement. We have also shown, on real data, that an actual leak on the same data given by \cite{vrachimis_2020_3902046} can be, on average, detected to the parameters of the competition (mean distance from the leak $260\ \mathrm{m}$ over 30 experiments). 

Even a single moving sensor, without stationary support, can provide a sizeable benefit for narrowing down leak location. Our results show that an iterative algorithm based on a hydraulic simulation can be an effective tool for leak localization with just a few, or possibly even one more sensor in addition to the current monitoring scheme. This result may be of great importance to water delivery companies in areas with fewer options for installing expensive sensor infrastructure.

Our contributions in this work are: 
\begin{enumerate}
	\item A novel, two-stage approach for leak localization with a minimal number of pressure sensors, including one that is mobile and can be moved between nodes. This is a significant improvement compared to standard methods -- like \cite{li2022leakage} (20 sensors for a network of less than 500 nodes) -- or even small sensor numbers methods as \cite{nejjari2015optimal} (where 5 sensors were used but for a network of 331 nodes) or \cite{righetti2019optimal} using 4 sensors but on a network of 23 users.
	\item In-depth discussion about the number of iterations for such a system to achieve convergence.
\end{enumerate}

\section{Iterative leak localization algorithm}\label{sec:method}

The description of the method used in this article is divided into two parts. First, we describe the basic algorithm of leak localization using simulations and applying a similarity measure. Then we present its development into the proposed iterative scheme.

\subsection{Simulations}
For this approach, we use the WDS hydraulic model (in our case, EPANET 2.2, \cite{rossman2000epanet}). The model is built with real-world data, including pipes' and nodes' geographical positions, elevation,  valve types, pipe material, etc., and measured nodes' water demand. The model allows us to simulate the WDS's behaviour with various parameters -- including node demands, state of the valves, leaks, and changed inflow to the WDS.

In our case, the simulations will be used to approximate the pressure values in nodes containing pressure sensors if the known leak is inserted in various places of the WDS.

\subsection{Basic algorithm for leak localization}

The algorithm for localizing the already detected leak presented in this Section represents a general approach to identifying the node where the leak is located using hydraulic simulations and pressure sensors. It is widely used in the field both for leak localization, as presented in papers such as \cite{farley2010field},  \cite{farley2013localization} \cite{jensen2018leakage}, or for sensor placement as in  \cite{steffelbauer2014sensor}, and it will also be the base of our iterative approach. 

\subsubsection{Preliminaries}

Let $\mathcal{M}$ be a WDS of $M$ points. Let then $\mathcal{N} \in \mathcal{M}$ be a subset of $N < M$ points where pressure sensors are located. Measurements on those sensors create $N \times T$ matrix pressure values, where $T$ is the length of set of time indices $\mathcal{T}$.

Let $P^z \in \mathbb{R}^{\mathcal{N} \times \mathcal{T}}$ be the measured pressure matrix, where $P^z[n,t] = p^{z}_{n,t}$ is the physical pressure measured in pressure sensor  $n\in\mathcal{N}$ at time $t\in\mathcal{T}$.

Let $D \in  \mathbb{R}^{\mathcal{M} \times \mathcal{T}}$ be demand matrix, where $D[m,t] = d_{m,t}$ be a water demand in node $m\in\mathcal{M}$ at time $t\in\mathcal{T}$. Let also $h \in \mathbb{R}^\mathcal{T}$ be the vector of pressures at the inflow of the WDS at times $\mathcal{T}$. Let also $D_{l_v, m}$ be a demand matrix modified by adding value $l_v$ to each element of row $m$.

Let $S: \mathbb{R}^{\mathcal{M} \times \mathcal{T}} \times \mathbb{R} \rightarrow \mathbb{R}^{\mathcal{N} \times \mathcal{T}}$ be a physical simulation function, that, given demand matrix and head pressure generates pressure matrix $P^{sw} \in \mathbb{R}^{\mathcal{N} \times \mathcal{T}}$ of simulated pressures at every node in $\mathcal{N}$.
	
Now, let $P^{sw}_{m}(l_v) =  P^{sw}_m \in \mathbb{R}^{\mathcal{N} \times \mathcal{T}}$, where $P^{sw}_m[n, t] = p^{sw}_{n,t,m}$ pressure simulated in sensor  $n\in\mathcal{N}$ at time $t\in\mathcal{T}$ by simulation $S(D_{l_v, m}, h)$.

\subsubsection{Input}

Input for the basic algorithm for leak localization is as follows:

\begin{itemize}
	\item A WDS of $M$ points
	\item Matrix $P^z$ of pressured measured in the WDS in time $\mathcal{T}$.
	\item Matrix $D$ of demands  in the WDS in time $\mathcal{T}$.
	\item A vector $h$ of head pressure values in the WDS in time $\mathcal{T}$.
	\item A real number $l_v$ of estimated leak size.
\end{itemize}

\subsubsection{Step 1: Calculating pressure matrices}
The first step of main method, as it is presented by authors of e.g \cite{steffelbauer:hal-03517648} is calculating the pressure matrices $P^{sw}_m$ for every $m\in \mathcal{N}$. These matrices are the main tools for describing differences in pressures between a normally working grid and the leaky one. 

\subsubsection{Step 2: Calculating measures of similarity}

The key element of using sensitivity matrix is to compare the results of simulation with leak placed in a candidate node with the data from the real grid. The use of sensitivity matrix is based on hypothesis of increased similarity between simulation and real data as the candidate node gets closer to actual leak. For measuring that, the similarity measure is used. It can vary, depending on the conditions of the particular network, from calculating changes from reference data (state of the network without the leak, real or simulated) to straightforward finding similarities between $P^z$ and $P^{sw}_m$ for $m\in\mathcal{M}$.

In method presented in this article, we propose to locate the node $\hat{m}$ as the one with the lowest square root of mean square error (\textit{RMSE}), as proposed by \cite{steffelbauer2022pressure}, of difference between pressures $p^z_{n,t}$ and $p^{sw}_{n, t, m}$, as follows:
\begin{equation}
	\hat{m}=\underset{m\in\mathcal{M}}{\operatorname{argmin}}\;RMSE(P^z,P^{sw}_m)
\end{equation}
where
\begin{equation}
	RMSE(P^z,P^{sw}_m) = \sqrt{\frac{1}{NT}\sum_{n=1}^N\sum_{t=1}^T(p^z_{n,t} - p^{sw}_{n, t, m})^2}.
\end{equation}

This measure is straightforward to use and since it is easy both to implement and analyse we have decided it forms a good base for the analysis of iterative framework we propose. 

We have decided to use this particular measure of similarity since the preliminary experiments showed that it proved better than other tested methods in the environment with very small amount of sensor data.

\subsubsection{Result}

The result of a basic algorithm for leak localization is node $\hat{m}$, for which, when a leak is added in a simulation, it produces simulated pressures $P^{sw}_m$ closest to the measured pressures $P^z$. Therefore, $\hat{m}$ is the node that the algorithm points out as the leak node.

\subsection{Proposed extension to the basic algorithm with iterative use of pressure matrix}

The proposed algorithm is a two-step approach to leak localization. Its core motivation is that there are a scarce number of sensors in many water networks, making precise localization very challenging. In such a situation, relocating the sensors when the approximate position of the leak is determined might significantly improve the effectiveness of the localization procedure.

The approach works under the assumption that we have $n$ sensors located in the section of the water grid and one additional sensor that can be moved from one node to another.
3
\subsubsection{First iteration} \label{ss:finidng}
Having real-life data from the actual WDS and the WDS's simulation model the single iteration of the algorithm performs as follows:
\begin{enumerate}
	\item Measure the real values of pressures in the WDS to obtain $P^z$,
	\item Proceed with $M$ simulations placing in each one the leak in different node $m$, to obtain matrices $P^{sw}_m$,
	\item For each simulation with leak located at $m \in \mathcal{M}$  calculate the $RMSE(P^z, P^{sw}_m)$. Those results are sorted in increasing order creating a list $R$ of ranks of each $m \in \mathcal{M}$.
	\item Select $m_s = \text{argmin}_{m \in \mathcal{M}} RMSE(P^z, P^{sw}_m) = R_0$, first element in the list $R$.
\end{enumerate}

The result of this step is the node $m_s$, for which adding a leak in the simulation yields the most similar results to measured data.

\subsubsection{The shift}
The second step of the algorithm is shifting the mobile sensor to the new available location. To achieve this, after first step, when node $m_s$ has been determined as the one most likely containing a leak according to first iteration:
\begin{enumerate}\label{def:iterative}
	\item If $m_s$ has not had a sensor placed in it  -- place the sensor at node $m_s$.
	\item If $m_s$ has had a sensor placed in -- place the sensor at the node closest to $m_s$. 
\end{enumerate}
This step is designed to move the mobile sensor closer to the leak location to achieve higher precision in leak localization.

\subsubsection{Subsequent iteration}
In this step the procedure of approximating leak's localization as described in \ref{ss:finidng} is repeated and a final result of the algorithm is obtained.

\section{Results}\label{sec:experiments}

To test the presented algorithm, we perform two experiments, first to test it on simulated data and finally applying it to the real leak detection problem. We use the L-Town network and leak data, as published in the BattLeDIM competition dataset~\cite{dataset}. This dataset is used as a benchmark as it is well-documented and accessible. It also contains the large number of nodes required to perform experiments testing the effectiveness of our two-stage algorithm.

\subsection{Dataset description}\label{sec:dataset}

The BattLeDIM dataset uses SCADA measurements of flow and pressure sensors installed within a water distribution system designated as L-Town, with $782$ nodes. It consists of a network hydraulic model, demand and leak patterns for the nodes in the network and actual measurement data taken during the time of the leaks in $33$ pre-determined locations. It provides a suitable environment for both simulating the behaviour of the L-Town networks as well as leakage patterns to simulate leaks in various nodes. The full description of both the dataset and BattLeDIM competition can be found in \cite{vrachimis_2020_3902046}.

The BattLeDIM dataset was utilized for testing various approaches towards leak detection, including the one presented in \cite{steffelbauer2022pressure} in the format of competition of methods results of which are presented in \cite{battledimresults}. It can be viewed as a benchmark for leak detection tasks. It also provides measurement data, making it a suitable environment for both simulated and real-life experiments. 

\subsection{Experiments}

To test our approach, we consider $n$ as the number of sensors, of which $n-1$ are stationary sensors and one is a mobile sensor that can be relocated between nodes. Let also $\mathcal{N}$ be the set of nodes on which to place the mobile sensor.

As a model leak, we use a leak of the size of $s = 6.38m^3/h$, which is around $3 \%$ of general use of the entire WDS and is the value of an actual leak measured there. To simulate the real data from the WDS, we will use noised simulations (a simulation made on the data where both demand information and leak pattern are influenced by Gaussian noise. That noise is also applied to output data of the simulations), the way proposed in \cite{SOLDEVILA2017Bayesian}, \cite{cuguero2015uncertainty},  \cite{PONCE2012extended}  with $l$ as upper limit of Gaussian noise on the pressure readings. As nodes available for sensor placement $\mathcal{N}$, we will use only 33 locations of sensors in the BattleDIM scenario.

We have prepared two experiments: one to test the approach on simulated leaks and the other on a real-life situation.

\paragraph{Experiment 1} The schema of this experiment is as follows: a simulated leak of the size $s = 6.38\ \frac{\mathrm{m}^3}{\mathrm{h}}$ is inserted into a node $m$. A noised simulation is used to simulate the real data, with $l = 0.1$. Then a $n$ initial sensors positions are selected at random, of which 1 is considered to be mobile one. The proposed algorithm will be then used to localize the leak. 

This experiment will be repeated for every node $m$ in WDS, for each of them $10$ times -- with randomly chosen $n$ initial sensors. That way the leak localization will be chosen for each node and in various selection of placement of sensors.

With $m=782$ and 10 repetitions for each node, that gives total of 7820 runs of the experiment I. The repetition of it for each node for varying choice of initial sensor placement allows also to take into consideration various detectability in different zones of diverse and comprehensive L-Town dataset and ensures statistically sound results.

\paragraph{Experiment 2} The schema of the experiment is as follows: we use the real data from the leak used in BattleDIM competition, localized on pipe p257 with hourly water loss at $6.38\ \frac{\mathrm{m}^3}{\mathrm{h}}$. The real-life measured data from the WDS (from \cite{vrachimis_2020_3902046}) are used as a data from measurement. We then randomly choose sensors positions and apply our approach to localize the leak. This experiment is repeated $30$ times with various starting sensors localizations.

\subsection{Results}

\subsubsection{Experiment 1}
Results are presented in Table~\ref{tab:results_main}. The Table presents, for subsequent iterations of setups (different number of sensors), performance characteristics (leak detection distance $D_{leak}$, distance from the leak to closest sensor $D_{sensor}$ and  rank of the location on the list of candidate location). Those results are averaged over 7820 repetitions (10 random starting positions of sensors for a leak in every node in examined WDS).

In all experiments, the first iteration, i.e. shifting a mobile sensor as close as possible to the most probable leak location, improved the result, reducing the distance to the leak node by 45-61\% and reducing the rank of the leak such that for three sensors and more it was on average in the top 12 most likely nodes. For only three sensors (2 stationary, one mobile), the average distance from the leak is~$\sim$180m, well below the distance of 300m adopted as the cut-off value in the performance estimation during BattLeDIM.

Even a small number of additional (stationary) sensors clearly impacts the algorithm performance, i.e., lower detection distance and leak rank. For a very small number of sensors (1-2), the second iteration improves the detection, but when 3 or more sensors are available, this is not the case. This is because the first iteration already places the sensor in the optimal location. 

\begin{figure*}
	\centering
	\begin{subfigure}[b]{0.32\linewidth}
		\includegraphics[width=\linewidth]{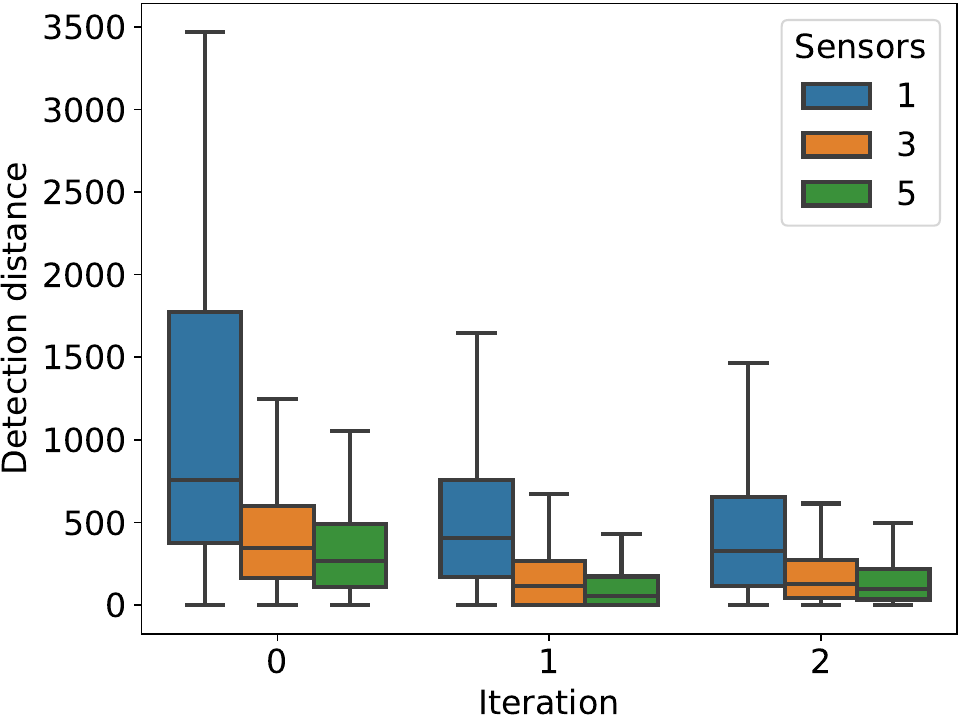}
		\caption{Detection distance}
	\end{subfigure}
	\begin{subfigure}[b]{0.32\linewidth}
		\includegraphics[width=1.0\linewidth]{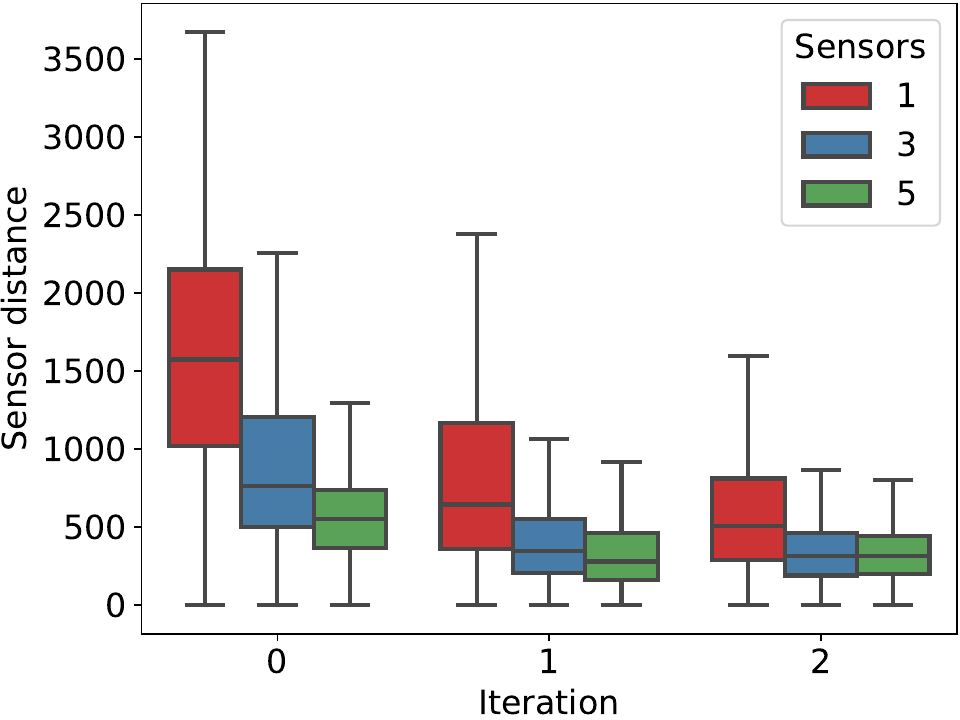}
		\caption{Sensor distance}
	\end{subfigure}
	\begin{subfigure}[b]{0.32\linewidth}
		\includegraphics[width=\linewidth]{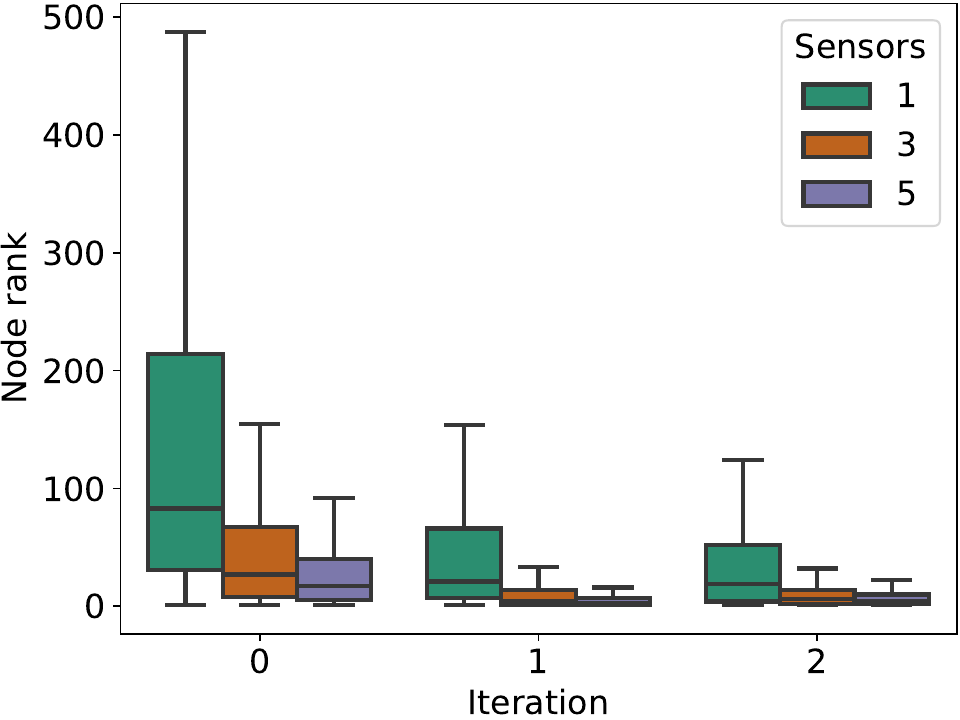}
		\caption{Node rank}
	\end{subfigure}
	\caption{The impact of the number of available sensors and algorithm iteration on its performance measures: detection distance, distance of the closes sensor to the leak node, node rank.}
	\label{fig:boxplots}
\end{figure*}

\begin{figure*}
	\centering
	\begin{subfigure}[b]{0.32\linewidth}
		\includegraphics[width=\linewidth]{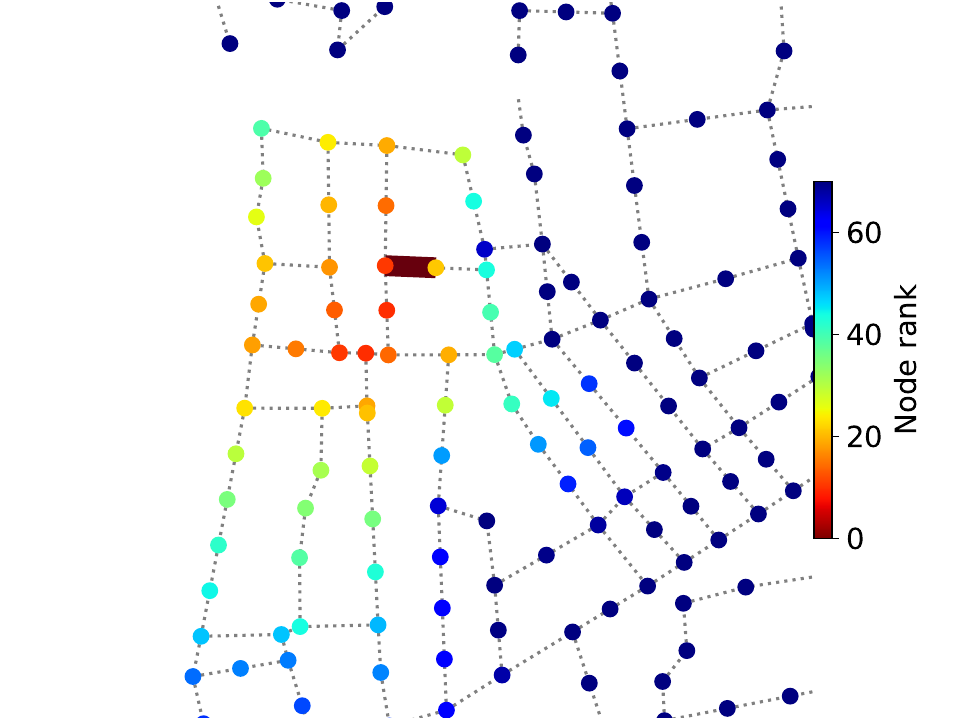}
		\label{fig:sample457}
	\end{subfigure}
	\begin{subfigure}[b]{0.32\linewidth}
		\includegraphics[width=\linewidth]{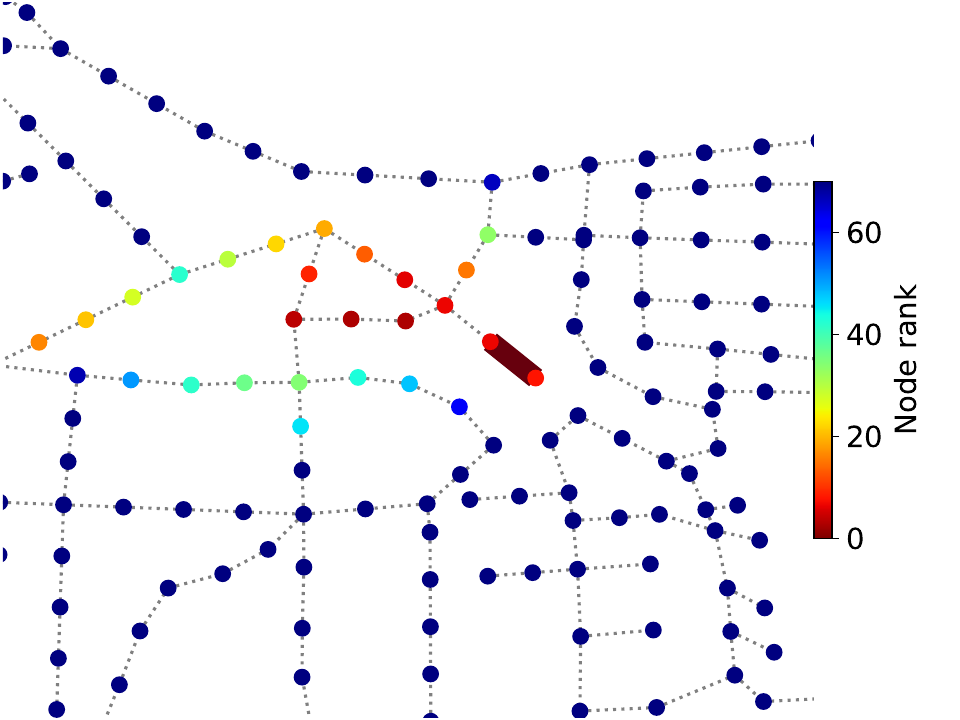}
		\label{fig:sample896}
	\end{subfigure}
	\begin{subfigure}[b]{0.32\linewidth}
		\includegraphics[width=\linewidth]{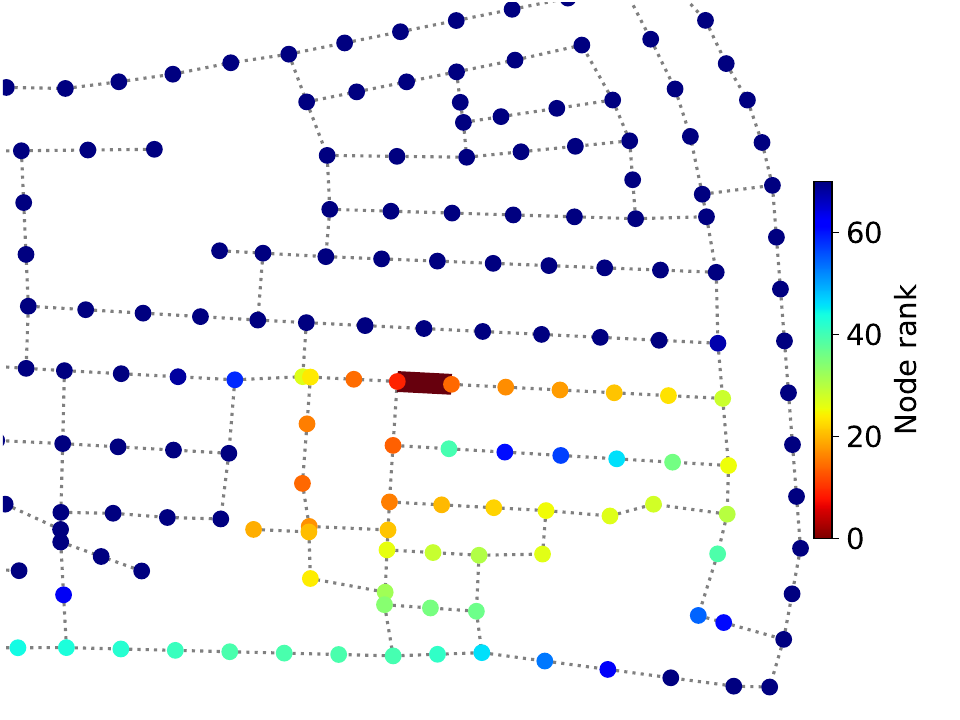}
		\label{fig:sample843}
	\end{subfigure}
	\caption{Examples of simulated leak localisation for links:  p457, p896, p843. Colours represent average node ranks over 5 iterations with different placement of three sensors (one mobile, two stationary).}
	\label{fig:samples}
\end{figure*}

Visualisation of the results is presented in Figure~\ref{fig:boxplots}. One of the main factors responsible for the high variance of the presented performance scores is the physical location of a leak node. Figure\ref{fig:graph_distance} presents the detection distance plotted on the DMA graph. Leaks are visibly more challenging to localise in some fragments of the graph. This difficulty may result from sparsely distributed locations of the possible sensor placement or large pipe diameters, making pressure drops associated with a leak more challenging to observe. Three example leak localisations in nodes adjacent to link p457, p896, and p843 are presented in~Figure~\ref{fig:samples}. Colours correspond to the average node rank over five repetitions of the experiment with different sensor placements, highlighting the low rank of the adjacent nodes, which corresponds to accurate leak localisation.

\noindent
\begin{minipage}{0.49\textwidth} 
	\centering 
	\vspace{15pt}
	\captionof{table}{Results of leak localisation Experiment 1 (simulated data). N$_s$ denotes the number of sensors, Iterations refers to shifts of the mobile sensor in consecutive days, D$_\text{leak}$ is the average detection distance and D$_\text{sens}$ is the average distance of the closest sensor to the leak. Rank refers to the ranking of the nodes suspected of being a leak location relative to the error measure of the algorithm. The values $\rho_\text{leak}, \rho_\text{sensor}, \rho_\text{rank}$ being standard deviation of those values. Distances are expressed as graph distances, values are averaged over five repetitions of the experiment. The line N$_s =33$ is a reference case where only one iteration is performed, but with sensors placed in all of 33 possible locations.}
	\label{tab:results_main}
	\small
	\begin{tabular}{rrrrrrrr}
		\FL
		N$_s$ & I & D$_\text{leak}$ & $\rho_\text{leak}$ & D$_\text{sensor}$ & $\rho_\text{sensor}$ & Rank & $\rho_\text{rank}$ \ML
		\multirow{3}{*}{1} & 0 & 1061.7 & 873.3 & 1612.2 & 768.4 & 168.0 & 194.6 \NN
		& 1 & 579.1 & 599.8 & 889.3 & 736.2 & 53.6 & 77.0 \NN
		& 2 & 454.7 & 450.4 & 656.4 & 554.3 & 37.8 & 59.1 \ML
		\multirow{3}{*}{3} & 0 & 424.9 & 353.5 & 900.0 & 551.1 & 40.4 & 38.2 \NN
		& 1 & 179.9 & 232.7 & 414.3 & 293.1 & 11.6 & 18.2 \NN
		& 2 & 189.5 & 229.5 & 336.6 & 195.4 & 10.5 & 13.1 \ML
		\multirow{3}{*}{5} & 0 & 343.7 & 318.4 & 566.3 & 278.1 & 25.4 & 24.5 \NN
		& 1 & 119.6 & 192.4 & 333.2 & 231.0 & 6.4 & 9.4 \NN
		& 2 & 149.8 & 201.6 & 327.1 & 172.9 & 7.9 & 12.7 \ML
		\multirow{3}{*}{9} & 0 & 231.4 & 243.0 & 409.1 & 216.4 & 13.9 & 15.5 \NN
		& 1 & 88.5 & 174.2 & 245.2 & 153.8 & 4.5 & 7.2 \NN
		& 2 & 158.0 & 214.9 & 317.7 & 165.0 & 8.0 & 13.5 \ML
		33 & - & 89.1 & 170.8 & 188.3 & 107.7 & 4.0 & 5.2 \LL
	\end{tabular}
\vspace{15pt}
\end{minipage}

\subsubsection{Experiment 2}
\begin{figure*}
	\centering
	\begin{subfigure}[b]{0.32\linewidth}
		\includegraphics[width=\linewidth]{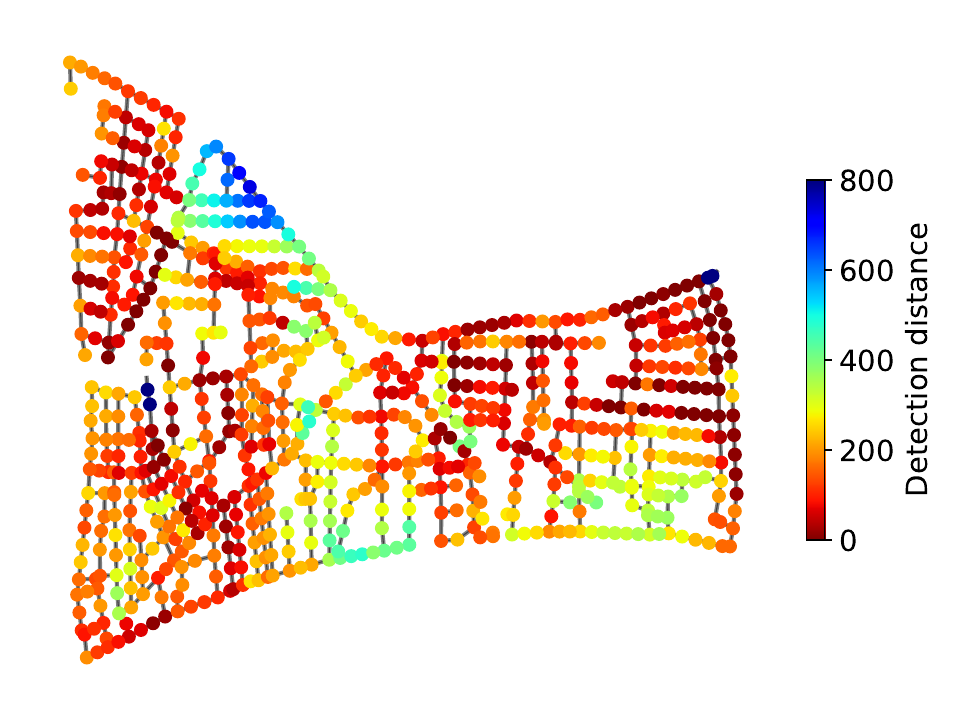}
		\caption{Average detection distance for all nodes in the DMA: first iteration of the algorithm, three sensors (one mobile, two stationary), five repetitions of the experiment with random initial sensor placement. Note the presence of `difficult' zones, where the detection distance exceeds~$300$m.}
		\label{fig:graph_distance}
	\end{subfigure}
	\begin{subfigure}[b]{0.32\linewidth}
		\includegraphics[width=\linewidth]{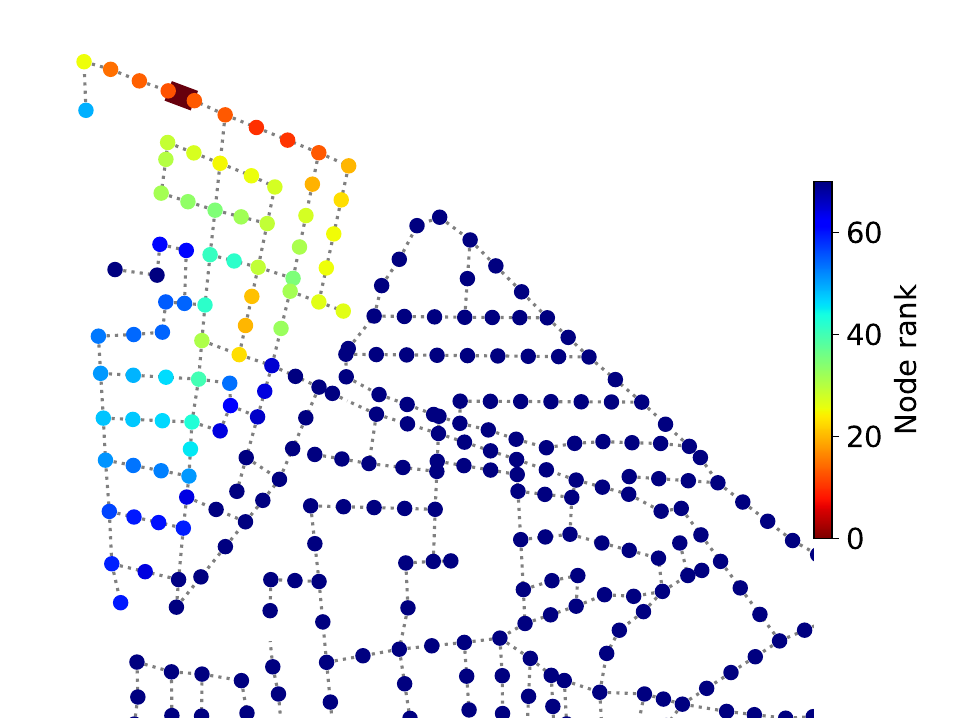}
		\caption{Visualisation of leak localisation using real data. The leak pipe in the NW corner is marked red. Colours represent average node ranks over 30 iterations with different placement of three sensors (one mobile, two stationary).}
		\label{fig:real_experiment}
	\end{subfigure}
	\begin{subfigure}[b]{0.32\linewidth}
		\includegraphics[width=\linewidth]{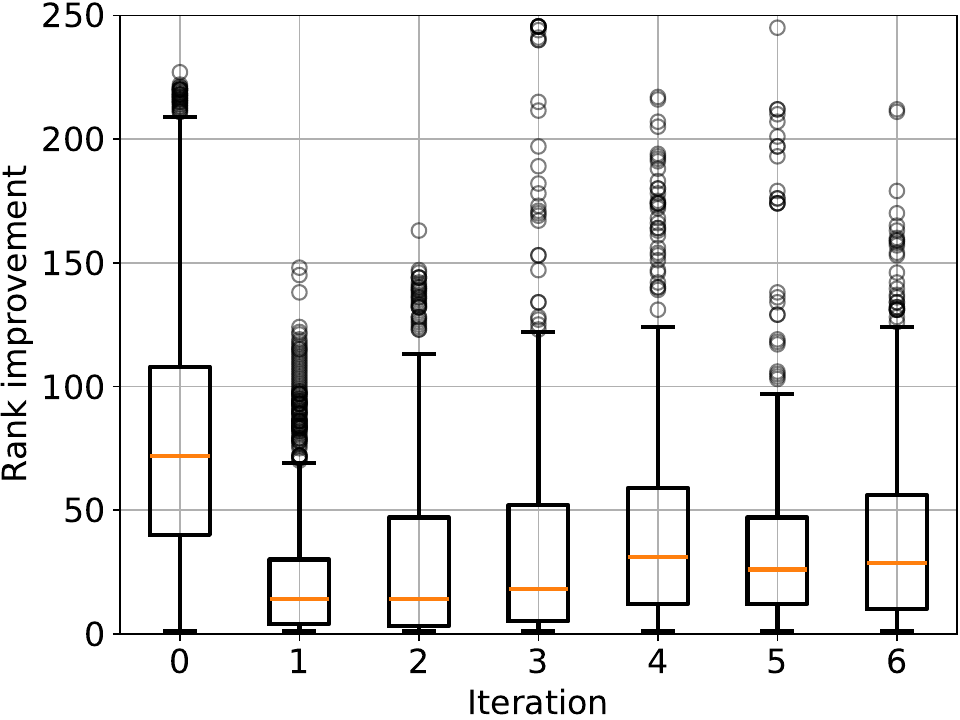}
		\caption{The rank of correct node for leak detection given to it by subsequent iterations of algorithm for leak localization if it had more than two stages. We can see how the rank achieves minimum in the second stage, then worsens. That results might be due to insufficient number of better nodes where the sensor can be placed.}
		\label{fig:more_ter}
	\end{subfigure}
	\caption{Visualisation of the leak detection/localisation results.}
	\label{fig:add}
\end{figure*}

Table.~\ref{tab:results_real} presents a summary of a leak localisation experiment performed on an actual leak data:  the leak on pipe \textit{p257} with hourly consumption of approx $6.38\ \frac{\mathrm{m}^3}{\mathrm{h}}$ that persisted for a long time and was present in 2018-01-22 where the experiment starts. In thirty experiments with three randomly placed sensors, the leak was localised on average with 13th rank and detection distance of 260m in the first iteration of the algorithm.  Fig.~\ref{fig:real_experiment} presents the experiment results as an average rank of the area where the leak is located; the most probable nodes are localised on the same street. This corresponds well with simulation results presented in Fig.~\ref{fig:graph_distance}, where the detection distance for nodes in this particular street is similar. 
\noindent 
\begin{minipage}{0.5\textwidth} 
	\centering 
	\vspace{15pt}
	\captionof{table}{Result of the leak localisation Experiment 2 (real data). N$_s$ denotes the number of used sensors, Iteration refers to shifts of the mobile sensor in consecutive days, D$_\text{leak}$ is the average detection distance and D$_\text{sensor}$ is the average distance of the closest sensor to the leak. Rank refers to the ranking of the nodes suspected of being a leak location relative to the error measure of the algorithm. The values $\rho_\text{leak}$, $\rho_\text{sensor}$, $\rho_\text{rank}$ being standard deviation of those values. Distances are expressed as graph distances, values are averaged over thirty repetitions of the experiment. The line N$_s =33$ is a reference case where only one iteration is completed, but with sensors placed in all of 33 possible locations.}
	\label{tab:results_real}
	\small
	\begin{tabular}{rrrrrrrr}
		\FL
		N$_s$ & I & D$_\text{leak}$ & $\rho_\text{leak}$ & D$_\text{sensor}$ & $\rho_\text{sensor}$ & Rank & $\rho_\text{rank}$ \ML
		\multirow{3}{*}{1} & 0 & 308.9 & 221.6 & 1585.7 & 834.3 & 47.5 & 28.6 \NN
		& 1 & 193.9 & 209.3 & 368.4 & 152.4 & 12.4 & 18.3 \NN
		& 2 & 306.8 & 127.7 & 537.6 & 113.8 & 9.0 & 4.6 \ML
		\multirow{3}{*}{3} & 0 & 423.3 & 249.3 & 1068.9 & 596.6 & 57.1 & 29.4 \NN
		& 1 & 259.2 & 200.9 & 431.5 & 158.1 & 12.9 & 15.6 \NN
		& 2 & 208.7 & 176.9 & 445.8 & 159.5 & 6.6 & 5.5 \ML
		\multirow{3}{*}{9} & 0 & 400.2 & 279.7 & 642.2 & 304.5 & 52.7 & 31.2 \NN
		& 1 & 194.8 & 179.7 & 393.8 & 162.1 & 8.1 & 9.5 \NN
		& 2 & 196.4 & 178.2 & 426.5 & 153.7 & 6.0 & 5.6 \ML
		33 & 0 & 0.0 & 0.0 & 255.0 & 0.0 & 1.0 & 0.0 \LL

	\end{tabular}
\vspace{15pt}
\end{minipage}

\begin{figure*}
	\centering
	\begin{subfigure}[b]{0.32\linewidth}
		\includegraphics[width=\linewidth]{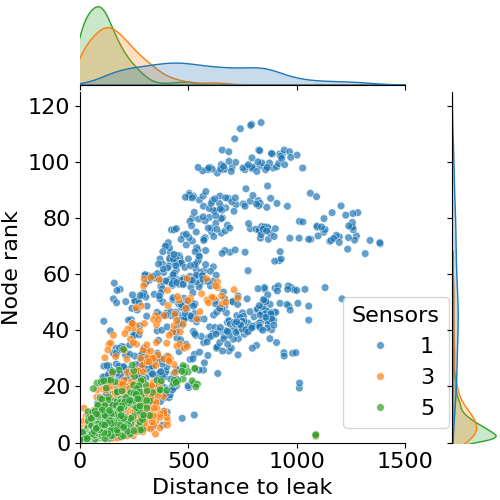}
		\caption{}
		\label{fig:scatter_a}
	\end{subfigure}
	\begin{subfigure}[b]{0.32\linewidth}
		\includegraphics[width=\linewidth]{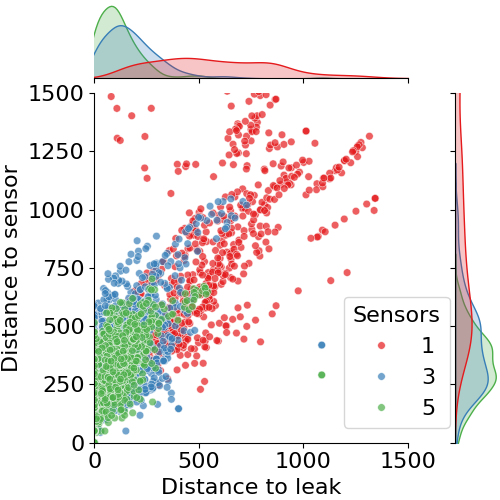}
		\caption{}
		\label{fig:scatter_b}
	\end{subfigure}
	\begin{subfigure}[b]{0.32\linewidth}
		\includegraphics[width=\linewidth]{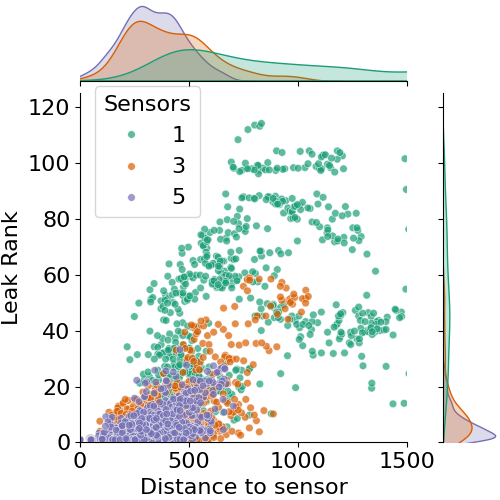}
		\caption{}
		\label{fig:scatter_c}
	\end{subfigure}
	\caption{Visualisation of the relationships between leak detection distance, leak node rank and distance of the leak to the nearest sensor. Every dot represents averaged result for five repetitions of the experiment with different initial position of sensors.}
	\label{fig:scatter}
\end{figure*}

\section{Discussion}

The results of both experiments show very promising effectiveness of the presented method -- when compared to reference experiment~\ref{tab:results_main}, having 33 sensors in the WDS,  the simulated experiment reaches very similar results using only nine sensors if 1st iteration and the result expected by BattLeDIM scenario are met with barely three, or even with one in Experiment II depicting real-life scenario (this particular leak is well detectable with only one sensor, on average the first iteration assigns the node 13th rank with detection distance 193m.).

A strong relationship between leak node rank and detection distance is presented in Fig.~\ref{fig:scatter_a}. The algorithm assigns a low rank to nodes close to the actual leak. A  relationship between the detection distance (between the node indicated as leaky and the actual leak) and the distance to the closest sensor to the leak is presented in Fig.~\ref{fig:scatter_b}. Intuitively, the proximity of a sensor helps in detecting leaks. This relationship is visible in Fig.~\ref{fig:scatter_c}; the distance between the nearest sensor and the leak node is inversely proportional to the leak node rank. However, a significant variance can be observed, especially for a single sensor case. A small number of sensors may require more than a single iteration.

All plots in Fig.~\ref{fig:boxplots} present a sharp, positive impact of the number of sensors on all performance measures. Even the presence of one or two stationary sensors significantly improves the results.

What is also interesting is that, excluding one sensor scenario, the correct node where the leak is located is presented in the top $2\%$ of the nodes. Comparing this result to Fig.~\ref{fig:scatter_a}, presenting the relation between the node pointed by the algorithm and the actual leak, it also shows that the algorithm, on average, shows the close neighbourhood of the actual leak -- usually, however, having problems with distinguishing which leak in the correct neighbourhood is the solution.

It is worth noticing that the results were achieved with only one mobile sensor and  33 nodes in the network available for sensor placement, which means that the possibility of sensor relocation was minimal in the network of almost 800 nodes altogether. The authors believe that increasing the number of available locations and the number of sensors that can be moved can improve those results. The obtained results are on par with the result given by \cite{marzola2022leakage} authors, of which used 33 sensors to obtain them, or \cite{li2022leakage} where a network of 491 nodes was used and 20 pressure sensors. A  method using the closest number of sensors to find the leak is described in \cite{righetti2019optimal}, but it covers a very small network of 23 users and 34 pipes.

Looking at those results, it might be enticing to continue to venture beyond the limit of two stages and iterate the same procedure to improve the results further. The preliminary experiments, however, show that -- at least with current parameters -- this is not the case. We can see the signs of that in Fig.~\ref{fig:boxplots}, showing that the second iteration does not yield significant improvement over the first when the number of sensors is three or more, and in time, that does not change -- and preliminary experiments performed for more iterations (see Fig.~\ref{fig:more_ter}) seem to show the same effect. The main reason for that is relatively good placement of sensors in the second stage of the algorithm and limited number of nodes where it can be placed better. It is possible that this could be improved with more nodes available for sensor placement, but further experimentation is required.

The results for only one iteration and a small number of sensors, however, confirm the theory that the tradeoff between the number of sensors and the time of leak detection is, in fact, possible. This may give water delivery companies in areas with less developed sensor infrastructure more options for leak localization.

\section{Conclusions}\label{sec:conclusions}

In this article, we have presented a two-stage schema for leak localization using a small number of pressure sensors, which maintains the effectiveness of state-of-the-art algorithms but at the cost of working time. This schema is less helpful Within the networks equipped with high-quality measure sensors as many methods utilize multi-sensor data well. In many cases, though, where the large section of the grid is limited to $2-3$ available sensors, the presented schema handles the tasks well, accurate and robust to usage and pressure noises generated by micro-leaks within the network.

This research is, however, to be further expanded. The most obvious choice from this point forward is to improve the quality of a single iteration of the algorithm, which can improve the overall efficiency in leak detection and possibly expand the accuracy improvement past the second iteration. Also, an interesting path is within the analysis of the availability of locations for installing the mobile sensor and its influence on the scheme's effectiveness. Another thing would be working on other limitations of old or less equipped water WDNs, such as lower pressure in pipes originating in larger diameters of the pipes in older systems.

\section*{Acknowledgements}
This work has been partially supported by the Polish National Centre for Research and Development grant  POIR.01.01.01-00-1414/20-00, `Intelligence Augumentation Ecosystem for analysts of water distribution networks'.

\bibliographystyle{plain}
\bibliography{Leak_detection_mascots}

\begin{thebibliography}{10}

\bibitem{avk}
10 ways to reduce water loss.
\newblock
  \url{https://www.avkvalves.eu/en/insights/water-technology/10-ways-to-reduce-water-loss}.
\newblock Accessed: 2023-11-07.

\bibitem{hawle}
How can you reduce non-revenue water levels?
\newblock
  \url{https://www.hawle.com/en/hawle-knowledge/basics/non-revenue-water/}.
\newblock Accessed: 2023-11-07.

\bibitem{diehl}
Non-revenue water.
\newblock
  \url{https://www.diehl.com/metering/en/products-solutions/metering-solutions/water-management-solutions/non-revenue-water/}.
\newblock Accessed: 2023-11-07.

\bibitem{Rastburg2020GamePlacement}
Georg Arbesser-Rastburg and Daniela Fuchs-Hanusch.
\newblock Serious sensor placement—optimal sensor placement as a serious
  game.
\newblock {\em Water}, 12(1), 2020.

\bibitem{casillas2013optimal}
Myrna~V Casillas, Vicen{\c{c}} Puig, Luis~E Garza-Castan{\'o}n, and Albert
  Rosich.
\newblock Optimal sensor placement for leak location in water distribution
  networks using genetic algorithms.
\newblock {\em Sensors}, 13(11):14984--15005, 2013.

\bibitem{cuguero2015uncertainty}
Pep Cuguer{\'o}-Escofet, Joaquim Blesa, Ramon P{\'e}rez, Miquel~A.
  Cuguero-Escofet, and Gerard Sanz.
\newblock Assessment of a leak localization algorithm in water networks under
  demand uncertainty.
\newblock {\em IFAC-PapersOnLine}, 48(21):226--231, 2015.
\newblock 9th IFAC Symposium on Fault Detection, Supervision andSafety for
  Technical Processes SAFEPROCESS 2015.

\bibitem{farah2017leakage}
Elias Farah and Isam Shahrour.
\newblock Leakage detection using smart water system: combination of water
  balance and automated minimum night flow.
\newblock {\em Water Resources Management}, 31(15):4821--4833, 2017.

\bibitem{farley2013localization}
B.~Farley, S.~R. Mounce, and J.~B. Boxall.
\newblock Development and field validation of a burst localization methodology.
\newblock {\em Journal of Water Resources Planning and Management},
  139(6):604--613, 2013.

\bibitem{farley2010field}
B~Farley, SR~Mounce, and JB~Boxall.
\newblock Field testing of an optimal sensor placement methodology for event
  detection in an urban water distribution network.
\newblock {\em Urban Water Journal}, 7(6):345--356, 2010.

\bibitem{glomb2023detection}
P~G{\l}omb, M~Cholewa, W~Koral, A~Madej, and M~Romaszewski.
\newblock Detection of emergent leaks using machine learning approaches.
\newblock {\em Water Supply}, 2023.

\bibitem{hamilton2013leak}
Stuart Hamilton and Bambos Charalambous.
\newblock {\em Leak detection: technology and implementation}.
\newblock IWA Publishing, 2013.

\bibitem{ISHIDO2014pressure}
Y.~Ishido and S.~Takahashi.
\newblock A new indicator for real-time leak detection in water distribution
  networks: Design and simulation validation.
\newblock {\em Procedia Engineering}, 89:411--417, 2014.
\newblock 16th Water Distribution System Analysis Conference, WDSA2014.

\bibitem{jensen2018leakage}
Tom~N{\o}rgaard Jensen, Vicen{\c{c}} Puig, Juli Romera, Carsten~Skovmose
  Kalles{\o}e, Rafa{\l} Wisniewski, and Jan~Dimon Bendtsen.
\newblock Leakage localization in water distribution using data-driven models
  and sensitivity analysis.
\newblock {\em Ifac-papersonline}, 51(24):736--741, 2018.

\bibitem{KAFLE2022acoustic}
Marshal~Deep Kafle, Stanley Fong, and Sriram Narasimhan.
\newblock Active acoustic leak detection and localization in a plastic pipe
  using time delay estimation.
\newblock {\em Applied Acoustics}, 187:108482, 2022.

\bibitem{khulief2012acoustic}
YA~Khulief, A~Khalifa, R~Ben Mansour, and MA~Habib.
\newblock Acoustic detection of leaks in water pipelines using measurements
  inside pipe.
\newblock {\em Journal of Pipeline Systems Engineering and Practice},
  3(2):47--54, 2012.

\bibitem{li2022leakage}
Xin Li, Shipeng Chu, Tuqiao Zhang, Tingchao Yu, and Yu~Shao.
\newblock Leakage localization using pressure sensors and spatial clustering in
  water distribution systems.
\newblock {\em Water Supply}, 22(1):1020--1034, 2022.

\bibitem{liemberger2006challenge}
Roland Liemberger, Philippe Marin, and B~Kingdom.
\newblock The challenge of reducing non-revenue water in developing
  countries--how the private sector can help: A look at performance-based
  service contracting.
\newblock 2006.

\bibitem{liemberger2019quantifying}
Roland Liemberger and Alan Wyatt.
\newblock Quantifying the global non-revenue water problem.
\newblock {\em Water Supply}, 19(3):831--837, 2019.

\bibitem{marzola2022leakage}
Irene Marzola, Filippo Mazzoni, Stefano Alvisi, and Marco Franchini.
\newblock Leakage detection and localization in a water distribution network
  through comparison of observed and simulated pressure data.
\newblock {\em Journal of Water Resources Planning and Management},
  148(1):04021096, 2022.

\bibitem{mashhadi2021use}
Neda Mashhadi, Isam Shahrour, Nivine Attoue, Jamal El~Khattabi, and Ammar
  Aljer.
\newblock Use of machine learning for leak detection and localization in water
  distribution systems.
\newblock {\em Smart Cities}, 4(4):1293--1315, 2021.

\bibitem{needs2021progress}
ACCELERATION NEEDS and F~Indicator.
\newblock Progress on level of water stress.
\newblock {\em Progress on the Level of Water Stress}, 2021.

\bibitem{nejjari2015optimal}
Fatiha Nejjari, Ramon Sarrate, and Joaquim Blesa.
\newblock Optimal pressure sensor placement in water distribution networks
  minimizing leak location uncertainty.
\newblock {\em Procedia Engineering}, 119:953--962, 2015.

\bibitem{neves2010evolvable}
Pedro Neves and Mauro Onori.
\newblock Evolvable production systems: Approach towards modern production
  systems.
\newblock In {\em Proceedings of the 6th CIRP-Sponsored International
  Conference on Digital Enterprise Technology}, pages 813--822. Springer, 2010.

\bibitem{PONCE2012extended}
Myrna Violeta~Casillas Ponce, Luis Eduardo~Garza Casta{\'n}{\'o}n, and
  Vicenc~Puig Cayuela.
\newblock Extended-horizon analysis of pressure sensitivities for leak
  detection in water distribution networks.
\newblock {\em IFAC Proceedings Volumes}, 45(20):570--575, 2012.
\newblock 8th IFAC Symposium on Fault Detection, Supervision and Safety of
  Technical Processes.

\bibitem{RICORAMIREZ2007StochasticPlacement}
Vicente Rico-Ramirez, Sergio Frausto-Hernandez, Urmila~M. Diwekar, and Salvador
  Hernandez-Castro.
\newblock Water networks security: A two-stage mixed-integer stochastic program
  for sensor placement under uncertainty.
\newblock {\em Computers \& Chemical Engineering}, 31(5):565--573, 2007.
\newblock ESCAPE-15.

\bibitem{righetti2019optimal}
Maurizio Righetti, Carlos Maximiliano~Giorgio Bort, Michele Bottazzi, Andrea
  Menapace, and Ariele Zanfei.
\newblock Optimal selection and monitoring of nodes aimed at supporting
  leakages identification in wds.
\newblock {\em Water}, 11(3):629, 2019.

\bibitem{rossman2000epanet}
Lewis~A Rossman et~al.
\newblock Epanet 2: users manual, 2000.

\bibitem{SOLDEVILA2022Relocation}
Adrià Soldevila, Joaquim Blesa, Sebastian Tornil-Sin, Rosa~M. Fernandez-Canti,
  and Vicen{\c{c}} Puig.
\newblock Incremental upgrading sensor placement methodology: Application to
  the leak localization in water networks.
\newblock {\em Computers \& Chemical Engineering}, 158:107642, 2022.

\bibitem{SOLDEVILA2017Bayesian}
Adrià Soldevila, Rosa~M. Fernandez-Canti, Joaquim Blesa, Sebastian Tornil-Sin,
  and Vicen{\c{c}} Puig.
\newblock Leak localization in water distribution networks using bayesian
  classifiers.
\newblock {\em Journal of Process Control}, 55:1--9, 2017.

\bibitem{steffelbauer:hal-03517648}
David Steffelbauer, Jochen Deuerlein, Denis Gilbert, Edo Abraham, and Olivier
  Piller.
\newblock {Pressure-Leak Duality for Leak Detection and Localization in Water
  Distribution Systems}.
\newblock {\em {Journal of Water Resources Planning and Management}}, 148(3),
  March 2022.

\bibitem{steffelbauer2014sensor}
David Steffelbauer, Markus Neumayer, Markus G{\"u}nther, and Daniela
  Fuchs-Hanusch.
\newblock Sensor placement and leakage localization considering demand
  uncertainties.
\newblock {\em Procedia Engineering}, 89:1160--1167, 2014.

\bibitem{steffelbauer2022pressure}
David~B Steffelbauer, Jochen Deuerlein, Denis Gilbert, Edo Abraham, and Olivier
  Piller.
\newblock Pressure-leak duality for leak detection and localization in water
  distribution systems.
\newblock {\em Journal of Water Resources Planning and Management},
  148(3):04021106, 2022.

\bibitem{vrachimis_2020_3902046}
Vrachimis, Eliades, Taormina, Ostfeld, Kapelan, Liu, Kyriakou, Pavlou, Qiu, and
  Polycarpou.
\newblock {BattLeDIM 2020 Problem Announcement and Description}, February 2020.

\bibitem{battledimresults}
Stelios Vrachimis, Demetrios Eliades, Riccardo Taormina, Zoran Kapelan, Avi
  Ostfeld, Shuming Liu, Marios Kyriakou, Pavlos Pavlou, Mengning Qiu, and
  Marios Polycarpou.
\newblock Battle of the leakage detection and isolation methods.
\newblock {\em Journal of Water Resources Planning and Management},
  148:04022068, 05 2022.

\bibitem{dataset}
Stelios~G. Vrachimis, Demetrios~G. Eliades, Riccardo Taormina, Avi Ostfeld,
  Zoran Kapelan, Shuming Liu, Marios~S. Kyriakou, Pavlos Pavlou, Mengning Qiu,
  and Marios Polycarpou.
\newblock {Dataset of BattLeDIM: Battle of the Leakage Detection and Isolation
  Methods}, September 2020.

\bibitem{yu2021integrated}
Jie Yu, Li~Zhang, Jinyu Chen, Yao Xiao, Dibo Hou, Pingjie Huang, Guangxin
  Zhang, and Hongjian Zhang.
\newblock An integrated bottom-up approach for leak detection in water
  distribution networks based on assessing parameters of water balance model.
\newblock {\em Water}, 13(6):867, 2021.

\bibitem{zaman2020review}
Dina Zaman, Manoj~Kumar Tiwari, Ashok~Kumar Gupta, and Dhrubjyoti Sen.
\newblock A review of leakage detection strategies for pressurised pipeline in
  steady-state.
\newblock {\em Engineering Failure Analysis}, 109:104264, 2020.

\end{thebibliography}
\end{document}